\documentclass[preprint, showkeys, showpacs,APS]{revtex4}
\begin{document}
\title{Young's and shear moduli and Poisson's ratio for elastic media of high and middle symmetry}

\author{T. Paszkiewicz}
\email{tapasz@prz.edu.pl} 
\author{S. Wolski}
\affiliation{Chair
of Physics, Rzesz\'{o}w University of Technology, Al. Powsta\'{n}c\'{o}w. Warszawy 6,
PL-35-959 Rzesz\'{o}w Poland}

\begin{abstract}
Using bases of fourth rank tensorial bases of  $\left[[V^{2}]^{2}\right]$ symmetry elaborated by Walpole we obtained expressions for inverse of Young's modulus $E(\textbf{n})$, inverse of shear modulus  $G(\textbf{m},\textbf{n})$ and Poisson's ratio $\nu(\textbf{m},\textbf{n})$, which depend on components of the stiffness tensor $\textbf{S}$, on direction cosines of vectors $\textbf{n}$ of uniaxial load and the vector $\textbf{m}$ of lateral strain with crystalline symmetry axes. Crystalline media of high and medium symmetries are considered. Such representation yields decomposition of the above elastic characteristics to isotropic and anisotropic parts. Expressions for Poisson's coefficient are well suited for studying the property of auxeticity. 
\end{abstract}

\maketitle

\section{Introduction} 
\label{introd}
Recent interest in single crystal materials has made the understanding their elastic characteristics, i.e. the compliances $C_{ij}$ and stiffness coefficients $S_{ij} $ \cite{t-g-j-apl-ph-2}, the bulk and Young's moduli $E$, Poisson's ratio $\nu$ and the shear modulus $G$, increasingly important   \cite{t-g-j-apl-ph-1}-\cite{li}. The anisotropy of Poisson's ratio was studied in several papers \cite{g-s-proc-roy-soc}-\cite{tokmakova}. In particular there is fundamental and practical interest in materials with negative Poisson's ratio, called auxetics \cite{e-n-h-r}. Love \cite{love} presents a single example of cubic \textit{single crystal} pyrite as having a Poisson's ratio of -0.14. Lakes described the synthesis of an actual auxetic material and proposed a simple mechanism underlying the negative Poisson's ratio \cite{lakes}. Alderson and Evans fabricated microporous polyethylene having a negative Poisson's ratio \cite{alderson-evans}. In spite of general belief it was shown that many crystalline solids, among them metals, have a negative Poisson's ratio \cite{m-h-pr}-\cite{l-m}. Data were compiled from the literature by Baughman et al. and were analyzed to show negative Poisson's ratio to occur for stress in an oblique direction upon single crystals of cubic metals \cite{baughman'98}.

In our previous paper \cite{p-p-w}, we considered Poisson's ratio of all cubic materials stretched in $[001]$ direction and measured for $[100]$ lateral direction. We concluded that auxeticity is a rather common phenomenon. 

Studying auxeticity of cubic elastic media we noticed that the size of the auxetic part of the stability region depends on the direction $\textbf{n}$ of stretch and the direction $\textbf{m}$ of lateral strain. To verify this observation we derived general expressions for $E(\textbf{n})$, $\nu(\textbf{m},\textbf{n})$, and $G(\textbf{m},\textbf{n})$ for all stable cubic materials. 

Our derivation gives expressions which do not depend on the choice of Cartesian coordinate system. To obtain such formulas, we used fourth-rank tensorial bases introduced by Walpole \cite{walpole}. This approach is simpler than the method developed by Turley and Sines \cite{t-g-j-apl-ph-2,t-g-j-apl-ph-1}, and applied by Gunton and Saunders \cite{g-s-j-m-s,ch-f,g-s-proc-roy-soc}, Li \cite{li} and Jain and Verma \cite{j-v}. To obtain Poisson's ratio and Young's and shear moduli for an arbitrary direction of applied stress, these authors used the method of rotations by Eulerian angles. 
\section{Poisson's ratio, Young's modulus and shear modulus}
\label{sc:definitions}
Elastic properties of an anisotropic body are characterized by the elastic constants $C_{ijkl}$ or by the elastic compliance coefficients $S_{ijkl}$ relative to an orthogonal coordinate system. Consider any two specified orthogonal unit vectors $\textbf{n}$ and $\textbf{m}$ and three related characteristics of elastic properties of the considered body \cite{sir-sha,rychl}, namely the Poisson ratio $\nu(\textbf{m},\textbf{n})$  
\begin{equation}
\nu(\textbf{m},\textbf{n})= \frac{\varepsilon_{\textbf{m}}}{\varepsilon_{\textbf{n}}}=-\frac{m_{i}m_{j}S_{ijkl}n_{k}n_{l}}{n_{\alpha}n_{\beta}S_{\alpha \beta \gamma \delta}n_{\gamma}n_{\delta}}, 
\label{eq:poisson}
\end{equation}
Young's modulus $E(\textbf{n})$
\begin{equation}
\frac{1}{E(\textbf{n})}= n_{i}n_{j}S_{ijkl}n_{k}n_{l},
\label{eq:young}
\end{equation}
and the shear modulus $G(\textbf{m},\textbf{n})$
\begin{equation}
\frac{1}{4G(\textbf{m},\textbf{n})}=m_{i}n_{j}S_{ijkl}m_{k}n_{l}. 
\label{eq:shear}
\end{equation}
Repetition of a suffix in a product of tensors or in a single tensor implies the usual summation with the respect to that suffix over the values 1,2,3.

Introduce three $\textbf{n}\otimes \textbf{n}$ and $\textbf{m}\otimes \textbf{m}$ and $\textbf{m}\otimes\textbf{n}$ with elements $n_{i}n_{j}$,  $m_{i}m_{j}$ and $m_{i}n_{j}$ ($i,j=1,2,3$). Using these dyads the definitions (\ref{eq:poisson})-(\ref{eq:shear}) can be written in a brief form (cf. Rychlewski \cite{rychl}). 
\begin{equation}
E(\textbf{n})^{-1}=(\textbf{n}\otimes \textbf{n})\cdot \textbf{S} \cdot (\textbf{n}\otimes \textbf{n}),
\label{eq:d-young}
\end{equation}
\begin{equation}
[4G(\textbf{m},\textbf{n})]^{-1}=[4G(\textbf{n},\textbf{m})]^{-1}=(\textbf{m}\otimes \textbf{n})\cdot \textbf{S} \cdot (\textbf{m}\otimes \textbf{n}),
\label{eq:d-shear}
\end{equation}
\begin{equation}
\frac{\nu(\textbf{m},\textbf{n})}{E(\textbf{n})}=\frac{\nu(\textbf{n},\textbf{m})}{E(\textbf{m})}=-(\textbf{m}\otimes \textbf{m})\cdot \textbf{S} \cdot (\textbf{n}\otimes \textbf{n}).
\label{eq:d-poisson}
\end{equation}
In an arbitrary chosen coordinate system, both $E^{-1}(\textbf{n})$ and $[4G(\textbf{m},\textbf{n})]^{-1}$ are related to quadratic forms of a positive definite matrix, hence $E^{-1}(\textbf{n})>0$, $[4G(\textbf{m},\textbf{n})]^{-1}>0$. 

Tensors $\textbf{C}$ and $\textbf{S}$ with elements $C_{ijkl}$ and $S_{ij,kl}$, respectively, obey the relation 
\begin{equation}
\textbf{C}\cdot \textbf{S}=\textbf{S}\cdot \textbf{C}=\textbf{I}_{4},
\label{eq:c times s}
\end{equation}
or for components
\begin{equation}
C_{ijrs}S_{rskl}=S_{ijrs}C_{rskl}=\frac{1}{2}\left(\delta_{ik}\delta_{jl}- \delta_{il}\delta_{jk}\right).
\label{eq:c times s-comp}
\end{equation}
This means that $\textbf{S}=\textbf{C}^{-1}$ and. 
\section{Poisson's ratio, Young's modulus and shear modulus for isotropic elastic media}
\label{sc:isotropic}
Introduce two fourth rank tensors -- $\textbf{J}$ with components 
\begin{equation}
J_{ijkl}=\frac{1}{3}\left(\textbf{I}_{2}\right)_{ij}\left(\textbf{I}_{2}\right)_{kl},
\nonumber 
\end{equation}
and $\textbf{K}=\left(\textbf{I}_{4}-\textbf{J}\right)$ (cf. Walpole \cite{walpole}). The second rank tensor $\textbf{I}_{2}$ has components $\left(I_{2}\right)_{ij}=\delta_{ij}$.

The introduced tensors obey several important relations 
\begin{eqnarray}
\textbf{J}^{2}=\textbf{J},\: \textbf{K}^{2}=\textbf{K},\nonumber \\
\textbf{J} + \textbf{K}=\textbf{I}_{4}, \: \textbf{J}\cdot\textbf{K}=\textbf{K}\cdot\textbf{J}=0. 
\label{eq:j and k}
\end{eqnarray}
The tensors $\textbf{J}$ and $\textbf{K}$ have complete Voigt symmetry. Using relations (\ref{eq:c times s})-(\ref{eq:j and k}), one obtains
\begin{eqnarray}
\textbf{C}=c_{J}\textbf{J}+c_{K}\textbf{K},\nonumber \\
\textbf{S}=s_{J}\textbf{J}+s_{K}\textbf{K}, 
\label{eq:expr-J and K}
\end{eqnarray}
where \cite{walpole} 
\begin{equation}
c_{J}=\frac{C_{iijj}}{3}=C_{11}+2C_{12},\: c_{K}=\frac{1}{5}\left(C_{ijij}-\frac{1}{3}C_{iijj}  \right)=\left(C_{11}-C_{12}\right),\: s_{J}=c_{J}^{-1},\:s_{K}=c_{K}^{-1}. 
\nonumber 
\end{equation}
This means that $\textbf{S}=c_{J}^{-1}\textbf{J}+ c_{K}^{-1}\textbf{K}$. An isotropic medium is mechanically stable if $c_{J}>0$, $c_{K}>0$ ($s_{J}>0$, $s_{K}>0$).

The tensor $\textbf{S}$ can be rewritten in a useful form 
\begin{equation}
	\textbf{S}=\left(s_{J}-s_{K}\right)\textbf{J}+s_{K}\textbf{I}_{4}.
	\label{eq:c-is-second-form}
\end{equation}

From the definition (\ref{eq:d-young}) and the form (\ref{eq:c-is-second-form}) of the tensor $\textbf{S}$, one obtains 
\begin{equation}
E^{-1}(\textbf{n})\equiv E^{-1}=\left(s_{J}-s_{K}\right)/3+s_{K}=s_{J}+2s_{K}/3>0.
\nonumber 
\end{equation}
Hence, using the tensorial basis we obtained the familiar expression for $E^{-1}(\textbf{n})$ \cite{sir-sha}, \cite{nye}. 

Calculating products $(\textbf{m}\otimes \textbf{m})\cdot \textbf{S} \cdot (\textbf{n}\otimes \textbf{n}$ and $(\textbf{m}\otimes \textbf{n})\cdot \textbf{S} \cdot (\textbf{m}\otimes \textbf{n})$, we obtain 
\begin{eqnarray}
\frac{\nu(\textbf{m},\textbf{n})}{E}=\frac{\left(s_{J}-s_{K}\right)}{3}=-\frac{S_{12}}{3}=\frac{C_{12}}{c_{J}c_{K}},\nonumber \\ 
\frac{1}{4G(\textbf{m},\textbf{n})}=\frac{s_{K}}{2}=\frac{1}{2c_{K}}>0. 
\nonumber 
\end{eqnarray} 
These formulas can be found in books \cite{sir-sha}, \cite{nye}. 

In agreement with Ting and Barnett \cite{ting-barnett}, an isotropic medium is completely auxetic (i.e. $\nu(\textbf{m},\textbf{n})<0$ for all pairs $(\textbf{m},\textbf{n})$), if $S_{12}>0$ (or $C_{12}<0$) and nonauxetic (i.e. $\nu(\textbf{m},\textbf{n})>0$ for all pairs $(\textbf{m},\textbf{n})$), if $S_{12}<0$ (or $C_{12}>0$). 
\section{Poisson's ratio, Young's modulus and shear modulus for cubic elastic media}
\label{sc:cubic}
As the three crystallographic directions of the cubic system are mutually perpendicular, the unit vectors $\textbf{a}$, $\textbf{b}$ and $\textbf{c}$ satisfy the conditions  
\begin{equation}
\textbf{a}\textbf{a}=\textbf{b}\textbf{b}=\textbf{c}\textbf{c}=1, \,	\textbf{a}\textbf{b}=\textbf{a}\textbf{c}=\textbf{b}\textbf{c}=0,\, \textbf{a}\otimes\textbf{a}+\textbf{b}\otimes\textbf{b}+\textbf{c}\otimes\textbf{c}=\textbf{I}_{2}. 
\nonumber 
\end{equation}
Introduce the symmetric fourth rank tensor $\Sigma$ with components 
\begin{equation}
\Sigma_{ijkl}=a_{i}a_{j}a_{k}a_{l}+b_{i}b_{j}b_{k}b_{l}+c_{i}c_{j}c_{k}c_{l}. 
\nonumber 
\end{equation}
With the help of $\Sigma$, Walpole \cite{walpole} constructed two different fourth rank tensors $\textbf{L}$ and $\textbf{M}$
\begin{equation}
\textbf{L}=\textbf{I}_{4}-\Sigma, \: \textbf{M}=\Sigma-\textbf{J}. 	
\nonumber 
\end{equation}
These tensors obey the following relations 
\begin{eqnarray}
\textbf{L}+\textbf{M}=\textbf{K}, \: \textbf{J}+\textbf{L}+\textbf{M}=\textbf{I}_{4},\nonumber \\
\textbf{L}^{2}=\textbf{L},\:\textbf{M}^{2}=\textbf{M},\:\nonumber \\
\textbf{J}\textbf{L}=\textbf{L}\textbf{J}=0,\: \textbf{J}\textbf{M}=\textbf{M}\textbf{J}=0,\: \textbf{L}\textbf{M}=\textbf{M}\textbf{L}=0. \nonumber 
\end{eqnarray}

For all cubic symmetry classes ($23$, $m3$, $432$, $\bar{4}3m$ and $m3m$) the tensors $\textbf{C}$, $\textbf{S}$ can be expanded in the basis of tensors $\textbf{J},\textbf{L},\textbf{M}$
\begin{eqnarray}
	\textbf{C}=c_{J}\textbf{J}+c_{L}\textbf{L}+c_{M}\textbf{M},\nonumber \\ 
	\textbf{S}=s_{J}\textbf{J}+s_{L}\textbf{L}+s_{M}\textbf{M},\label{eq:s-expanded}
\end{eqnarray}
where \cite{walpole}  
\begin{eqnarray}
c_{J}=C_{11}+2C_{12},\:	c_{L}=\frac{1}{3}L_{ijkl}C_{ijkl}=2C_{44},\: c_{M}=\frac{1}{2}M_{ijkl}C_{ijkl}=C_{11}-C_{12},\nonumber \\ 
s_{J}=c_{J}^{-1}, \: s_{L}=c_{L}^{-1}, \: s_{M}=c_{M}^{-1}. 
\nonumber 
\end{eqnarray}
$c_{J}$, $c_{L}$ and $c_{M}$ are the eigenvalues of the tensor $\textbf{C}$. Similarly, $s_{J}$, $s_{L}$ and $s_{M}$ are eigenvalues of $\textbf{S}$. The mechanical stability is guaranted when all these eigenvalues are positive.

The tensor $\textbf{S}$ (\ref{eq:s-expanded}) can be written in a useful form 
\begin{equation}
	\textbf{S}=\left(s_{J}-s_{M}\right)\textbf{J}+s_{L}\textbf{I}_{4}+\left(s_{M}-s_{L}\right)\Sigma.
	\label{eq:second-form-S}
\end{equation}

The coefficients $c_{J}$, $c_{L}$, and $c_{M}$ can be written in terms of $S_{11}$, $S_{12}$, and $S_{44}$, namely 
\begin{equation}
	s_{J}=S_{11}+2S_{12}, \, s_{L}=S_{44}/2,\, s_{M}=S_{11}-S_{12}.
\label{eq:w2v}
\end{equation}
Using these relations and second of Eqs. (\ref{eq:second-form-S}) we express the matrix elements $S_{ij}$ by $C_{ij}$ ($i,j=1,2,4$)
\begin{equation}
	S_{11}=\frac{C_{11}+C_{12}}{\left(C_{11}+2C_{12}\right)\left(C_{11}-C_{12}\right)},\, 	  		S_{12}=-\frac{C_{12}}{\left(C_{11}+2C_{12}\right)\left(C_{11}-C_{12}\right)},\, S_{44}=\frac{1}{C_{44}}.
\nonumber 
\end{equation}
These expressions are in agreement with familiar results \cite{nye}. 

We calculate Young's modulus in direction $\textbf{n}$ with the help of formulae (\ref{eq:d-poisson}) and (\ref{eq:second-form-S}). As a result, we obtain the familiar formulae \cite{nye}
\begin{eqnarray}
E^{-1}(\textbf{n})=\left[\left(s_{J}-s_{M}\right)/3+s_{L}\right]+\left(s_{M}-s_{L}\right)T(\textbf{n})=S_{11}-2\left(S_{11}-S_{12}-S_{44}/2\right)\nonumber \\
\times\left(n_{a}^{2}n_{b}^{2}+n_{b}^{2}n_{c}^{2}+n_{c}^{2}n_{a}^{2}\right),
\nonumber 
\end{eqnarray}
where $T\left(\textbf{n}\right)=\sum_{i=a}^{c}n_{i}^{4}$.

Similarly, for $\nu(\textbf{m},\textbf{n})$ and $G(\textbf{m},\textbf{n})$, we get
\begin{eqnarray}	
-\frac{\nu(\textbf{m},\textbf{n})}{E(\textbf{n})}=\left(s_{J}-s_{M}\right)/3+\left(s_{M}-s_{L}\right)P(\textbf{m},\textbf{n}),\label{nu-cub}\\ 		
\left[4G(\textbf{m},\textbf{n})\right]^{-1}=s_{L}/2+\left(s_{M}-s_{L}\right)P(\textbf{m},\textbf{n}). 
\nonumber 
\end{eqnarray}
The function $P\left(\textbf{m},\textbf{n}\right)$ was introduced by Ting and Barnett \cite{ting-barnett} 
\begin{equation}
P\left(\textbf{m},\textbf{n}\right)=P\left(\textbf{n},\textbf{m}\right) =\sum_{i=a}^{c}\left(m_{i}n_{i}\right)^{2},
\nonumber 
\end{equation}
where, for example, $m_{a}=(\textbf{m}\textbf{a})$. Note that the first term of expression defining $\left[-\nu(\textbf{m},\textbf{n})/E({\textbf{n}})\right]$ is isotropic, whereas the second one is anisotropic. The same structure has the shear modulus. 

The ratios $\nu /E$ and $[4G]^{-1}$ are not independent 
\begin{equation}
-\frac{\nu(\textbf{m},\textbf{n})}{E(\textbf{n})}-\frac{1}{4G(\textbf{m},\textbf{n})}=\frac{1}{3}\left(s_{J}-s_{M}\right)-\frac{1}{2}s_{L}.
\nonumber
\end{equation}
   
If $C_{44}=\left(C_{11}-C_{12}\right)/2$, one deals with an isotropic medium. In this case, $s_{L}=s_{M}\equiv s_{K}$, therefore the last terms of $E^{-1}(\textbf{n})$, $\nu(\textbf{m},\textbf{n})/E(\textbf{n})$ and $[4G(\textbf{m},\textbf{n})]^{-1}$ vanish, and one gets the results obtained in Sect. \ref{sc:isotropic}. 

Using Eq. (\ref{eq:w2v}) one can write $\nu(\textbf{m},\textbf{m})$ (\ref{nu-cub}) in the form obtained by Ting and Barnett \cite{ting-barnett}
\begin{equation}
	\nu(\textbf{m},\textbf{n})=\left[1-2P(\textbf{m},\textbf{n})\right]S_{12}+QP(\textbf{m},\textbf{n})/2,
\nonumber 
\end{equation}
where Q=$\left[2\left(S_{11}+S_{12}\right)-S_{44}\right]$.
The functions $T(\textbf{n})$, $P(\textbf{m},\textbf{n})$ obey the inequalities 
\begin{eqnarray}
	1/3\leq T\left(\textbf{n}\right)\leq 1, \nonumber\\ 
	0\leq P\left(\textbf{m},\textbf{n}\right)\leq 1/2. 
\nonumber 
\end{eqnarray}
The function $p(\textbf{n})$ achieves the minimal value for $\textbf{n}=\left\langle111\right\rangle$ and maximal value for $\textbf{n}=\left\langle001\right\rangle$. The function $P(\textbf{n},\textbf{m})$ attains the minimal value 0 for $\textbf{n}$ (or $\textbf{m}$) equal to $\left\langle001\right\rangle$ and for arbitrary unit vector $\textbf{m}$, perpendicular to this $\textbf{n}$ (or for arbitrary unit vector $\textbf{n}$ perpendicular to $\textbf{m}=\left\langle001\right\rangle$). The maximal value $1/2$ of $F(\textbf{n},\textbf{m})$ is reached for $\textbf{n}$ equal to $\left\langle-110\right\rangle$ or 
$\left\langle110\right\rangle$ and $\textbf{m}=\left\langle110\right\rangle$ or $\textbf{m}=\left\langle-110\right\rangle$.

The ratio $[-\nu(\textbf{m},\textbf{n})/E(\textbf{n})]$ obeys the inequality 
\begin{equation}
	S_{12}\leq -\nu(\textbf{m},\textbf{n})/E(\textbf{n})\leq Q/4.
\nonumber
\end{equation}
This means that in agreement with Ting and Barnett \cite{ting-barnett}, a cubic medium is completely auxetic if $S_{12}>0$ and $Q>0$, and nonauxetic if $S_{12}<0$, $Q<0$. If none of these pairs of inequalities hold, $\nu(\textbf{m},\textbf{n})$ is negative for some pairs $(\textbf{m},\textbf{n})$ and positive for other. 
\section{Characteristics of elasticity for transversely isotropic media}
\label{sc:trans-isotr}
Denote the direction of the symmetry axis by $\textbf{c}$ and a component of it by $c_{i}$ ($i=1,2,3$). Using the vector $\textbf{c}$, one can construct two basic \emph{second} rank tensors, namely a dyad $\textbf{p}=\textbf{c}\otimes\textbf{c}$ with components $\left(\textbf{p}\right)_{ij}\equiv p_{ij}=c_{i}c_{j}$ and $\textbf{q}=\left(\textbf{I}_{2}-\textbf{p}\right)$. Tensors $\textbf{p}$ and $\textbf{q}$ are idempotent, i.e. $\textbf{p}^{2}=\textbf{p}$, $\textbf{q}^{2}=\textbf{q}$. Their products vanish,  $\textbf{p}\cdot\textbf{q}=\textbf{q}\cdot\textbf{p}=0$.  

In the case of transversely isotropic media, the tensorial basis consists of five tensors $\textbf{E}_{1}$, $\textbf{E}_{2}$, $\textbf{E}_{s}$, $\textbf{F}$, and $\textbf{G}$ \cite{walpole}
\begin{eqnarray}
\left(\textbf{E}_{1}\right)_{ijkl}\equiv E^{(1)}_{ijkl}=p_{ij}p_{kl}, \: \left(\textbf{E}_{2}\right)_{ijkl}\equiv E^{(2)}_{ijkl}=q_{ij}q_{kl}/2,\nonumber \\ 	
	\textbf{G}_{ijkl}\equiv G_{ijkl}=\left(p_{ik}q_{jl}+p_{il}q_{jk}+p_{jk}q_{il}+p_{jl}q_{ik}\right)/2,
\nonumber 
\end{eqnarray}
\begin{equation}
	\left(\textbf{E}_{s}\right)_{ijkl}\equiv E^{(s)}_{ijkl}=\left[\left(\textbf{E}_{3}\right)_{ijkl}+\left(\textbf{E}_{4}\right)_{ijkl}\right]\equiv
	\left(p_{ij}q_{kl}+q_{ij}p_{kl}\right)/\sqrt{2},
	\nonumber 
\end{equation}
\begin{equation}
	\textbf{F}_{ijkl}\equiv F_{ijkl}=\left(q_{ik}q_{jl}+q_{jk}q_{il}-q_{ij}q_{kl}\right)/2.
	\label{eq:transv-isotr-F} 
\end{equation}
Tensors $\textbf{E}_{i}$ obey the multiplication table \ref{tab:E's}.
\begin{table}
		\centering
		\begin{tabular}
[c]{c|cccc}
\hline\noalign{\smallskip}
&$\mathbf{E}_{1}$ & $\mathbf{E}_{2}$ & $\mathbf{E}_{3}$ & $\mathbf{E}_{4}$\\
\noalign{\smallskip}\hline\noalign{\smallskip}
  $\mathbf{E}_{1}$ & $\mathbf{E}_{1}$ & $0$ & $\mathbf{E}_{3}$ & $0$\\
  $\mathbf{E}_{2}$ & $0$ & $\mathbf{E}_{2}$ & $0$ & $\mathbf{E}_{4}$\\
  $\mathbf{E}_{3}$ & $0$ & $\mathbf{E}_{3}$ & $0$ & $\mathbf{E}_{1}$\\
  $\mathbf{E}_{4}$ & $\mathbf{E}_{4}$ & $0$ & $\mathbf{E}_{2}$ & $0$
    \end{tabular}
	\caption{Multiplication table for $\textbf{E}$'s tensors}		
\label{tab:E's}
\end{table}

Similarly as $\textbf{E}_{1}$ and $\textbf{E}_{2}$, tensors $\textbf{F}$ and $\textbf{G}$ are idempotent, i.e. $\textbf{F}^{2}=\textbf{F}$, $\textbf{G}^{2}=\textbf{G}$ and 
\begin{equation}
	\textbf{F}\cdot\textbf{G}=\textbf{G}\cdot\textbf{F}=\textbf{0},\: \textbf{F}\cdot\textbf{E}_{\alpha}=\textbf{E}_{\alpha}\cdot\textbf{F}= \textbf{0}, \:  \textbf{G}\cdot\textbf{E}_{\alpha}=\textbf{E}_{\alpha}\cdot\textbf{G}= \textbf{0} \: (\alpha =1,\ldots 4).
	\label{eq:remaining-rel}
\end{equation}
For transversely isotropic media (symmetry classes $\infty$, $\infty /m$, $\infty 2$, $\infty m$ and $\infty /mm$ and $6$, $\bar{6}$, $6/m$, $622$, $6mm$, $\bar{6}m2$ and $6/mmm$), the tensor of elastic constants $\textbf{C}$ can be expanded in the introduced basis 
\begin{equation}
	\textbf{C}=c_{1}\textbf{E}_{1}+c_{2}\textbf{E}_{2}+c_{s}\textbf{E}_{s}+c_{F}\textbf{F}+c_{G}\textbf{G},
\nonumber 
\end{equation}
where \cite{walpole} 
\begin{eqnarray}
c_{1}=p_{ij}c_{ijkl}p_{kl}=C_{33},\: c_{2}=\frac{1}{2}q_{ij}C_{ijkl}q_{kl}=\left(C_{11}+C_{12}\right),\nonumber\\ 	
c_{s}=\frac{1}{\sqrt{2}}\left(p_{ij}C_{ijkl}q_{kl}+q_{ij}C_{ijkl}p_{kl}\right)=\sqrt{2}C_{13},\nonumber\\ 
c_{F}=\frac{1}{2}F_{ijkl}C_{ijkl}=\left(C_{11}-C_{12}\right), \:c_{G}=\frac{1}{2}G_{ijkl}C_{ijkl}=2C_{44}.
\label{eq:C-ti}
\end{eqnarray}
In the same way we find the coefficients of expansion of $\textbf{S}$ 
\begin{equation}
s_{1}=S_{33},\, s_{2}=(S_{11}+S_{12}),\, s_{s}=\sqrt{2}S_{13},\, s_{F}=\left(S_{11}-S_{12}\right),\, s_{G}=S_{44}/2.
\label{eq:coeff-S} 
\end{equation}

Since $\textbf{S}$ is inverse of $\textbf{C}$ and $\textbf{I}_{4}=\left(\textbf{E}_{1}+\textbf{E}_{2}+\textbf{F}+\textbf{G}\right)$, we find 
\begin{equation}
	\textbf{S}=s_{1}\textbf{E}_{1}+s_{2}\textbf{E}_{2}+s_{s}\textbf{E}_{s}+s_{F}\textbf{F}+s_{G}\textbf{G}\equiv \textbf{s}_{\bot},
	\label{eq:transv-iso-S}
\end{equation}
where
\begin{equation}
	s_{1}=\frac{c_{2}}{s_{E}}, \: s_{2}=\frac{c_{1}}{s_{E}},\: s_{s}=-\frac{c_{s}}{s_{E}},
	s_{F}=c_{F}^{-1}, \: s_{G}=c_{G}^{-1}, 
	\label{eq:ti-coef-S}
\end{equation}
where $s_{E}=\left(c_{1}c_{2}-c_{s}^{2}\right)$. These equations are in agreement with relation obtained by Boas and Mackenzie (cf. Nye \cite{nye}).

With the help of the expansion formulas \cite{walpole}
\begin{eqnarray}
	\textbf{J}=\frac{1}{3}\textbf{E}_{1}+\frac{2}{3}\textbf{E}_{2}+\frac{\sqrt{2}}{3}\textbf{E}_{s},\nonumber\\ 
	\textbf{K}=\frac{2}{3}\textbf{E}_{1}+\frac{1}{3}\textbf{E}_{2}-\frac{\sqrt{2}}{3}\textbf{E}_{s}+\textbf{F}+\textbf{G},
	\label{J-K-expansion}
\end{eqnarray}
and Eq. (\ref{eq:expr-J and K}), one can check that when $C_{44}=\left(C_{11}-C_{12}\right)/2$, $C_{33}=C_{11}$, $C_{13}=C_{12}$, transversely isotropic media are equivalent to isotropic media.  

Having the explicit form of expansion of $\textbf{S}$ tensor with the help of definitions (\ref{eq:d-young})-(\ref{eq:d-shear}) for transversely isotropic media we obtain 
\begin{eqnarray}
	E^{-1}(\textbf{n})=s_{1}\left(\textbf{n}\textbf{c}\right)^{4}+\frac{1}{2}\left(s_{2}+s_{F} \right)\left[1-\left(\textbf{n}\textbf{c}\right)^{2}    \right]^{2}+\left(\sqrt{2}s_{s}+2s_{G}\right)\left(\textbf{n}\textbf{c}\right)^{2}\left[1-\left(\textbf{n}\textbf{c}\right)^{2}\right],\nonumber\\ 
-\frac{\nu(\textbf{m},\textbf{n})}{E(\textbf{n})}=\left(s_{1}+s_{F}-2s_{G}\right)\left(\textbf{m}\textbf{c}\right)^{2}\left(\textbf{n}\textbf{c}\right)^{2}+\frac{1}{2}\left(s_{2}-s_{F}\right)\left[1-\left(\textbf{m}\textbf{c}\right)^{2}\right]\left[1-\left(\textbf{n}\textbf{c}\right)^{2}\right]\nonumber\\ 
+\frac{1}{\sqrt{2}}s_{s}\left\{\left(\textbf{m}\textbf{c}\right)^{2}\left[1-\left(\textbf{n}\textbf{c}\right)^{2}\right]+\left(\textbf{n}\textbf{c}\right)^{2}\left[1-\left(\textbf{m}\textbf{c}\right)^{2}\right]\right\},\nonumber\\ 
\label{eq:tr-isot-all}
\end{eqnarray}
\begin{eqnarray}
	\frac{1}{4G(\textbf{m}\textbf{n})}= \left(s_{1}+\frac{1}{2}s_{2}-\sqrt{2}s_{s}-s_{G}\right)\left(\textbf{m}\textbf{c}\right)^{2}\left(\textbf{n}\textbf{c}\right)^{2}+\frac{1}{2}s_{F}\left[1-\left(\textbf{m}\textbf{c}\right)^{2}\right]\left[1-\left(\textbf{n}\textbf{c}\right)^{2}\right]\nonumber\\ 
+\frac{1}{2}s_{G}\left\{\left(\textbf{m}\textbf{c}\right)^{2}\left[1-\left(\textbf{n}\textbf{c}\right)^{2}\right]+\left(\textbf{n}\textbf{c}\right)^{2}\left[1-\left(\textbf{m}\textbf{c}\right)^{2}\right]\right\}.	
\nonumber 
\end{eqnarray}
\section{Elastic characteristics of tetragonal elastic media}
\label{sc:tetragonal}
In the case of tetragonal elastic media, one deals with two groups of classes. To a more symmetrical belong $4mm$, $\bar{4}2m$, $422$ and $4/mmm$. The remaining less symmetric classes are $4$, $\bar{4}$ and $4/m$. 

As the crystallographic directions of all classes of the tetragonal system are mutually perpendicular, the related three unit vectors $\textbf{a}$, $\textbf{b}$, $\textbf{c}$ also are mutually perpendicular, hence
\begin{equation}
\textbf{a}\otimes\textbf{a}+\textbf{b}\otimes\textbf{b}+\textbf{c}\otimes\textbf{c}=\textbf{I}_{2}.
\label{eq:completness}
\end{equation}
Vector $\textbf{c}$ defines principal axis of symmetry. 
\subsection{Classes $4mm$, $\bar{4}2m$, $422$ and $4/mmm$}
\label{sc:tetr-more}
If vectors $\textbf{m}$ and $\textbf{n}$ are perpendicular, 
\begin{equation}
(\textbf{m}\textbf{a})(\textbf{n}\textbf{a})+(\textbf{m}\textbf{b})(\textbf{n}\textbf{b})+(\textbf{m}\textbf{c})(\textbf{n}\textbf{c})=0.
\label{eq:n-m-perp}
\end{equation}

Components of tensors $\textbf{E}_{1}$, $\textbf{E}_{2}$, $\textbf{E}_{s}$ and $\textbf{G}$ can be expressed by $\textbf{a}$, $\textbf{b}$ and $\textbf{c}$ \cite{walpole}
\begin{eqnarray}
	E^{(1)}_{ijkl}=c_{i}c_{j}c_{k}c_{l}, \: E^{(2)}_{ijkl}=\frac{1}{2}\left(a_{i}a_{j}+b_{i}b_{j}\right)\left(a_{k}a_{l}+b_{k}b_{l}\right),\nonumber\\  
E^{(3)}_{ijkl}=\frac{1}{\sqrt{2}}c_{i}c_{j}\left(a_{k}a_{l}+b_{k}b_{l}\right),\: E^{(4)}_{ijkl}=\frac{1}{\sqrt{2}}\left(a_{i}a_{j}+b_{i}b_{j}\right)c_{k}c_{l},\: \textbf{E}_{s}=\left(\textbf{E}_{3}+\textbf{E}_{4}\right),\nonumber\\ 
G_{ijkl}=\frac{1}{2}\left[\left(c_{i}a_{j}+a_{i}c_{j}\right)\left(c_{k}a_{l}+a_{k}c_{l}\right)+\left(c_{i}b_{j}+b_{i}c_{j}\right)\left(c_{k}b_{l}+b_{k}c_{l}\right)\right].
\nonumber 
\end{eqnarray}
Introduce two further tensors $\textbf{F}_{1}$ and $\textbf{F}_{2}$ with components
\begin{equation}
F^{(1)}_{ijkl}=\frac{1}{2}\left(a_{i}b_{j}+b_{i}a_{j}\right)\left(a_{k}b_{l}+b_{k}a_{l}\right), \: F^{(2)}_{ijkl}=\frac{1}{2}\left(a_{i}a_{j}-b_{i}b_{j}\right)\left(a_{k}a_{l}-b_{k}b_{l}\right).
	\label{eq:t1-Fs}
\end{equation}
Using the relation of completness (\ref{eq:completness}), one may show that $\left(\textbf{F}_{1}+\textbf{F}_{2}\right)=\textbf{F}$, where $\textbf{F}$ is defined by Eq. (\ref{eq:transv-isotr-F}). 

The introduced tensors make up the decomposition of $\textbf{I}_{4}$
\begin{equation}
\textbf{I}_{4}=\textbf{E}_{1}+\textbf{E}_{2}+\textbf{F}_{1}+\textbf{F}_{2}+\textbf{G}. 
	\label{eq:t1-decomp-unity}
\end{equation}
Tensors $\textbf{E}_{\alpha}$ ($\alpha =1,\ldots,4$) obey multiplication table \ref{tab:E's}. $\textbf{F}_{1}$, $\textbf{F}_{2}$, and $\textbf{G}$ are each idempotent, and $\textbf{F}_{1}\textbf{F}_{2}=\textbf{F}_{2}\textbf{F}_{1}=\textbf{0}$,   $\textbf{F}_{\beta}\textbf{G}=\textbf{G}\textbf{F}_{\beta}=\textbf{0}$ ($\beta =1,2$). Products of tensors $\textbf{F}_{\beta}$ and $\textbf{E}_{\alpha}$  vanish. 

As previously, both tensors $\textbf{C}$ and $\textbf{S}$ can represented in form of linear combinations \cite{walpole}
\begin{eqnarray}	\textbf{C}=c_{E}^{(1)}\textbf{E}_{(1)}+c_{E}^{(2)}\textbf{E}_{2}+c_{E}^{(s)}\textbf{E}_{s}+c^{(1)}_{F}\textbf{F}_{1}+c^{(2)}_{F}\textbf{F}_{2}+c_{G}\textbf{G},\\ \nonumber
\textbf{S}=s_{E}^{(1)}\textbf{E}_{1}+s_{E}^{(2)}\textbf{E}_{2}+s_{E}^{(s)}\textbf{E}_{s}+s^{(1)}_{F}\textbf{F}_{1}+s^{(2)}_{F}\textbf{F}_{2}+s_{G}\textbf{G},
	\label{eq:t1-C-S}
\end{eqnarray}
where $c_{E}^{(\alpha)}$ (previously we omitted the index E) and $c_{G}$ are the same as in the case of transversely isotropic media , whereas  
$s_{E}^{(\alpha)}$ $(\alpha =1,2,s)$  and $s_{G}$ obey Eqs. (\ref{eq:ti-coef-S}) . The coefficients $c^{(1)}_{F}$, $c^{(2)}_{F}$ and $s^{(1)}_{F}$, $s^{(2)}_{F}$ are equal  
\begin{equation}
	 c^{(1)}_{F}=2C_{66}, \: c^{(2)}_{F}=\left(C_{11}-C_{12}\right),\: 
	s^{(1)}_{F}=S_{66}/2=1/c^{(1)}_{F}, \: s^{(2)}_{F}=\left(S_{11}-S_{12}\right)=1/c^{(2)}_{F}. 
	\label{t1-F-c-s}
\end{equation}

If $S_{66}/2=\left(C_{11}-C_{12}\right)$, then $s_{F}^{(1)}=s_{F}^{(2)}=s_{F}$ and $s_{F}^{(1)}\textbf{F}_{1}+s_{F}^{(2)}\textbf{F}_{2}=s_{F}\textbf{F}$, hence $\textbf{s}_{t1}=\textbf{s}_{\bot}$. Analogous relations hold for $\textbf{C}$. 

Since the first of equations (\ref{J-K-expansion}) holds, and 
\begin{equation}
	\textbf{L}=\textbf{F}_{1}+\textbf{G},\; \textbf{M}=\frac{2}{3}\textbf{E}_{1}+\frac{1}{3}\textbf{E}_{2}-\frac{\sqrt{2}}{3}\textbf{E}_{s}+\textbf{F}_{2}
\nonumber
\end{equation}
when $S_{33}=S_{11}$, $S_{13}=S_{12}$, and $S_{66}=S_{44}$ the compliance tensors of cubic and tetragonal higher symmetry media coincide. 

The definitions of Poissson's ratio, Young's and shear moduli (\ref{eq:poisson})-(\ref{eq:shear}), and the relation (\ref{eq:n-m-perp}) in the case of tetragonal media of higher symmetry (labeled by the index t1), lead to the expressions
\begin{eqnarray}
-\frac{\nu_{t1}(\textbf{m},\textbf{n})}{E_{t1}(\textbf{n})}=\left[-\frac{\nu(\textbf{m},\textbf{n})}{E(\textbf{n})}\right]_{\bot}'+2s^{(1)}_{F}(\textbf{m}\textbf{a})(\textbf{m}\textbf{b})(\textbf{n}\textbf{a})(\textbf{n}\textbf{b})
 \nonumber\\
+\frac{1}{2}s^{(2)}_{F}\left[(\textbf{m}\textbf{a})^{2}-(\textbf{m}\textbf{b})^{2}\right]\left[(\textbf{n}\textbf{a})^{2}- (\textbf{n}\textbf{b})^{2}\right],
\nonumber 
\end{eqnarray}
\begin{equation}
E_{t1}^{-1}(\textbf{n})=\left[E^{-1}(\textbf{n})\right]_{\bot}'+2s^{(1)}_{F}(\textbf{n}\textbf{a})^{2}(\textbf{n}\textbf{b})^{2}
+\frac{1}{2}s^{(2)}_{F}\left[\left(\textbf{n}\textbf{a}\right)^{2}- \left(\textbf{n}\textbf{b}\right)^{2}\right]^{2},
\nonumber 
\end{equation}
\begin{eqnarray}
\frac{1}{4G_{t1}(\textbf{m},\textbf{n})}=\left[\frac{1}{4G(\textbf{m},\textbf{n})}\right]_{\bot}'+\frac{1}{2}s^{(1)}_{F}\left[\left(\textbf{m}\textbf{a}\right)\left(\textbf{n}\textbf{b}\right)+\left(\textbf{m}\textbf{b}\right)\left(\textbf{n}\textbf{a}\right)\right]^{2}\nonumber\\  +\frac{1}{2}s^{(2)}_{F}\left[\left(\textbf{m}\textbf{a}\right)\left(\textbf{n}\textbf{a}\right)-\left(\textbf{m}\textbf{b}\right)\left(\textbf{n}\textbf{b}\right)\right]^{2}, 
\nonumber 
\end{eqnarray}
where
\begin{eqnarray}
\left[-\frac{\nu(\textbf{m},\textbf{n})}{E(\textbf{n})}\right]_{\bot}'=\left(s_{1}-2s_{G}\right)\left(\textbf{m}\textbf{c}\right)^{2}\left(\textbf{n}\textbf{c}\right)^{2}+ \frac{1}{2}s_{2}\left[1-\left(\textbf{m}\textbf{c}\right)^{2}\right]\left[1-\left(\textbf{n}\textbf{c}\right)^{2}\right]\nonumber\\ 
+\frac{1}{\sqrt{2}}s_{s}\left\{\left(\textbf{m}\textbf{c}\right)^{2}\left[1-\left(\textbf{n}\textbf{c}\right)^{2}\right]+\left(\textbf{n}\textbf{c}\right)^{2}\left[1-\left(\textbf{m}\textbf{c}\right)^{2}\right]\right\},\nonumber\\ 
\left[E^{-1}(\textbf{n})\right]_{\bot}'=s_{1}\left(\textbf{n}\textbf{c}\right)^{4}+\frac{1}{2}s_{2}\left[1-\left(\textbf{n}\textbf{c}\right)^{2}    \right]^{2}+\left(\sqrt{2}s_{s}+2s_{G}\right)\left(\textbf{n}\textbf{c}\right)^{2}\left[1-\left(\textbf{n}\textbf{c}\right)^{2}\right],\nonumber\\ 
\left[\frac{1}{4G(\textbf{m},\textbf{n})}\right]_{\bot}'=\left(s_{1}+\frac{1}{2}s_{2}-\sqrt{2}s_{s}-s_{G}\right)\left(\textbf{m}\textbf{c}\right)^{2}\left(\textbf{n}\textbf{c}\right)^{2}\nonumber\\ 
+\frac{1}{2}s_{G}\left\{\left(\textbf{m}\textbf{c}\right)^{2}\left[1-\left(\textbf{n}\textbf{c}\right)^{2}\right]+\left(\textbf{n}\textbf{c}\right)^{2}\left[1-\left(\textbf{m}\textbf{c}\right)^{2}\right]\right\}.	
\nonumber
\end{eqnarray}
  
\subsection{Classes $4$, $\bar{4}$, $422$ and $4/m$}
\label{sc:tetr-less}
In the case of tetragonal less symmetric media (we shall label them by t2) to tensors $\textbf{E}_{\alpha}$ ($\alpha =1,\ldots,4$), $\textbf{F}_{1}$, $\textbf{F}_{2}$, and $\textbf{G}$, one should add two tensors $\textbf{F}_{3}$ and $\textbf{F}_{4}$
\begin{equation}
	\textbf{F}_{3}=\frac{1}{2}\left(\textbf{a}\otimes\textbf{b}+\textbf{b}\otimes\textbf{a}\right)\otimes\left(\textbf{a}\otimes\textbf{a}-\textbf{b}\otimes\textbf{b}\right), \:\textbf{F}_{4}=\frac{1}{2}\left(\textbf{a}\otimes\textbf{a}-\textbf{b}\otimes\textbf{b}\right) \otimes\left(\textbf{a}\otimes\textbf{b}+\textbf{b}\otimes\textbf{a}\right).
	\label{F-3-4}
\end{equation}
Together with tensors $\textbf{F}_{1}$, $\textbf{F}_{2}$ (\ref{eq:t1-Fs}), they form a subalgebra which is isomorphic to that of $\textbf{E}_{\alpha}$ ($\alpha =1,\ldots,4$) by having the multiplication table \ref{tab:E's} with the kernel letter $E$ replaced by $F$. The $\textbf{0}$ tensor is obtained when either of $\textbf{F}_{3}$ and $\textbf{F}_{4}$ multiplies (from the left or the right) $\textbf{G}$ or any one of $\textbf{E}$. 

As previously, the tensors $\textbf{C}$ and $\textbf{S}$ can be constructed as the linear combination 
\begin{eqnarray}
\textbf{C}=c_{1}\textbf{E}_{1}+c_{2}\textbf{E}_{2}+c_{s}\textbf{E}_{s}+c^{(1)}_{F}\textbf{F}_{1}+c^{(2)}_{F}\textbf{F}_{2}+c^{(s)}_{F}\textbf{F}_{s}+c_{G}\textbf{G},\\ \nonumber	\textbf{S}=s_{E}^{(1)}\textbf{E}_{1}+s_{E}^{(2)}\textbf{E}_{2}+s_{E}^{(s)}\textbf{E}_{E}^{(s)}+s^{(1)}_{F}\textbf{F}_{1}+s^{(2)}_{F}\textbf{F}_{2}+s^{(s)}_{F}\textbf{F}_{s}+s_{G}\textbf{G},
	\label{eq:t2-C-S}
\end{eqnarray}
where $\textbf{F}_{s}=\left(\textbf{F}_{3}+\textbf{F}_{4}\right)$, and  
\begin{equation}
	c^{(s)}_{F}=c^{(3)}_{F}=c^{(4)}_{F}=2C_{16},\, \; s^{(s)}_{F}=s^{(3)}_{F}=s^{(4)}_{F}=S_{16},
	\label{eq:C-Fs}
\end{equation}
$c_{E}^{(\alpha)}$ and $s_{E}^{(\alpha)} \; (\alpha=1,2,s)$  are given respectively by Eqs. (\ref{eq:C-ti}) and (\ref{eq:coeff-S})

In a similar manner to that explained previously, the scalar expressions 
\begin{equation}
	s^{(1)}_{F,t2}=\frac{c^{(2)}_{F}}{c^{(1)}_{F}c^{(2)}_{F}-\left[c^{(s)}_{F}\right]^{2}}, \: s^{(2)}_{F,t2}=\frac{c^{(1)}_{F}}{c^{(1)}_{F}c^{(2)}_{F}-\left[c^{(s)}_{F}\right]^{2}},\: s^{(s)}_{F,t2}=-\frac{c^{(s)}_{F}}{c^{(1)}_{F}c^{(2)}_{F}-\left[c^{(s)}_{F}\right]^{2}}, 
\nonumber 
\end{equation}
can be obtained. Using the explixit expressions for coefficients of expansions (\ref{eq:t2-C-S}), we get 
\begin{equation}
	S_{66}=\frac{C_{11}-C_{12}}{s_{F}},\; \left(S_{11}-S_{12}\right)=\frac{C_{66}}{s_{F}},\; S_{16}=-\frac{C_{16}}{s_{F}},
\nonumber %
\end{equation}
where $s_{F}=\left[\left(C_{11}-C_{12}\right)C_{66}-2C_{16}^{2}\right]$. Expressions relating $c_{E}^{(\alpha)}$ and $s_{E}^{(\alpha)}$, ($\alpha=1,2,s$) as well as $c_{G}$, and $s_{G}$, are the same as for t1 media.

Young's, shear, and the Poisson's coefficients for tetragonal media of lower symmetry are equal 
\begin{eqnarray}
\left[-\frac{\nu(\textbf{m},\textbf{n})}{E(\textbf{n})}\right]_{t2}=\left[-\frac{\nu(\textbf{m},\textbf{n})}{E(\textbf{n})}\right]_{t1}'+s^{(s)}_{F}\left\{(\textbf{m}\textbf{a})(\textbf{m}\textbf{b})\left[\left(\textbf{n}\textbf{a}\right)^{2}-\left(\textbf{n}\textbf{b}\right)^{2}\right]\right.\nonumber\\ 
\left.+(\textbf{n}\textbf{a})(\textbf{n}\textbf{b})\left[\left(\textbf{m}\textbf{a}\right)^{2}-\left(\textbf{m}\textbf{b}\right)^{2}\right]\right\},
\nonumber 
\end{eqnarray}
\begin{equation}
E^{-1}_{t2}(\textbf{n})=\left[E^{-1}(\textbf{n})\right]_{t1}'+2s^{(s)}_{F}\left(\textbf{n}\textbf{a}\right)\left(\textbf{n}\textbf{b}\right)^{2}\left[\left(\textbf{n}\textbf{a}\right)^{2}-\left(\textbf{n}\textbf{b}\right)^{2}\right],  
\nonumber 
\end{equation}
\begin{eqnarray}
\frac{1}{4G_{t2}(\textbf{m},\textbf{n})}=\left[\frac{1}{4G_{t1}(\textbf{m},\textbf{n})}\right]_{t1}'+ s^{(s)}_{F}\left[ \left(\textbf{m}\textbf{a}\right)\left(\textbf{n}\textbf{b}\right)+\left(\textbf{m}\textbf{b}\right)\left(\textbf{n}\textbf{a}\right)\right]\nonumber\\
\times\left[\left(\textbf{m}\textbf{a}\right)\left(\textbf{n}\textbf{a}\right)-\left(\textbf{m}\textbf{b}\right)\left(\textbf{n}\textbf{b}\right)\right],
\nonumber 
\end{eqnarray}
where $\left[-\frac{\nu(\textbf{m},\textbf{n})}{E(\textbf{n})}\right]_{t1}'$, $\left[E^{-1}(\textbf{n})\right]_{t1}'$, and $\left[\frac{1}{4G(\textbf{m},\textbf{n})}\right]_{t1}'$ can be obtained from $\left[-\frac{\nu_{t1}(\textbf{m},\textbf{n})}{E_{t1}(\textbf{n})}\right]$, $E^{-1}_{t1}(\textbf{n})$, and $\frac{1}{4G_{t1}(\textbf{m},\textbf{n})}$ as a result of the changes of coefficients $s_{F}^{(\alpha)}$ to $s_{F,t2}^{(\alpha)}$ ($\alpha =1,2$).

\section{Conclusions}
We demonstrated the effectivness of application of tensorial bases of $[[V^{2}]^{2}]$ for derivation the explicit expressions for Young's and shear moduli and for Poisson's ratio. These formulae are well suited for studying the anisotropy properties of $Y$, $G$, and $\nu$ \cite{p-w}. We considered materials of high and middle crystalline symmetry. Work on deriviation of the suitable expressions for low symmetry materials and 2D systems is currently in progress. 

\section*{Acknowledgements}We express our gratitude to R. Lakes for critical reading of the manuscript.

\end{document}